\documentclass[useAMS,usenatbib]{mn2e}
\usepackage{graphicx}
\usepackage{times}
\usepackage{wrapfig}
\makeatletter
\renewcommand{\thefootnote}{\fnsymbol{footnote}}

\newcommand{\Rmnum}[1]{\expandafter\@slowromancap\romannumeral #1@}
\makeatother

\title[UV spectral slope and obscuration]{GAMA/H-ATLAS: The ultraviolet spectral slope and obscuration in galaxies}

\author[D. B.Wijesinghe et al.]
{D. B. Wijesinghe$^{1}$\thanks{E-mail:D.Wijesinghe@physics.usyd.edu.au},
E. da Cunha$^{2}$, A. M. Hopkins$^{3}$, L. Dunne$^{4}$, R. Sharp$^{5}$,
\newauthor
M. Gunawardhana$^{1}$, S. Brough$^{3}$,
E. M. Sadler$^{1}$, S. Driver$^{6}$, I. Baldry$^{7}$,  S. Bamford$^{8}$, 
\newauthor
J. Liske$^{9}$, J. Loveday$^{10}$, P. Norberg$^{11}$, J. Peacock$^{11}$,  C. C. Popescu$^{12}$, R. Tuffs$^{13}$, E. Andrae$^{13}$,
\newauthor
R. Auld$^{14}$, M. Baes$^{15}$, J. Bland-Hawthorn$^{1}$, S. Buttiglione$^{16}$, A. Cava$^{17}$, E. Cameron$^{18}$,
\newauthor
C. J, Conselice$^{4}$, A. Cooray$^{19}$, S. Croom$^{1}$, A. Dariush$^{14,20}$, G. DeZotti$^{16}$, S. Dye$^{14}$, S. Eales$^{14}$,
\newauthor
C. Frenk$^{21}$, J. Fritz$^{22}$, D. Hill$^{6}$, R. Hopwood$^{23}$, E. Ibar$^{24}$, R. Ivison$^{24}$, M.Jarvis$^{25}$, D. H. Jones$^{3}$,
\newauthor
E. van Kampen$^{9}$, L. Kelvin$^{6}$, K. Kuijken$^{26}$, S. J. Maddox$^{4}$, B. Madore$^{27}$, M. J. Micha{\l}owski$^{11}$,
\newauthor
B. Nichol$^{28}$, H. Parkinson$^{11}$, E. Pascale$^{14}$, K. A. Pimbblet$^{29}$, M. Pohlen$^{14}$, M. Prescott$^{7}$,
\newauthor
G. Rodighiero$^{16}$, A. S. G. Robotham$^{6}$, E. E. Rigby$^{4}$, M. Seibert$^{26}$, S. Serjeant$^{23}$,
\newauthor
D. J. B. Smith$^{4}$, P. Temi$^{30}$, W. Sutherland$^{31}$, E. Taylor$^{1}$, D. Thomas$^{24}$, P. van der Werf$^{32}$ \\
$^{1}$Sydney Institute for Astronomy, School of Physics, University of Sydney, NSW 2006, Australia\\
$^{2}$Max-Planck Institute for Astronomy, Koenigstuhl 17, 69117 Heidelberg, Germany\\
$^{3}$Australian Astronomical Observatory, PO Box 296, Epping, NSW 1710, Australia\\
$^{4}$School of Physics \& Astronomy, University of Nottingham, University Park, Nottingham NG7 2RD, UK\\
$^{5}$Research School of Astronomy \& Astrophysics, Australian National University, Cotter Road, Weston Creek, ACT 2611, Australia\\
$^{6}$School of Physics \& Astronomy, University of St Andrews, North Haugh, St Andrews, KY16 9SS, UK\\
$^{7}$Astrophysics Research Institute, Liverpool John Moores University, Twelve Quays House, Egerton Wharf, Birkenhead, CH41 1LD, UK\\
$^{8}$Centre for Astronomy and Particle Theory, University of Nottingham, University Park, Nottingham, NG7 2RD, UK\\
$^{9}$European Southern Observatory, Karl-Schwarzschild-Str.~2, 85748, Garching, Germany\\
$^{10}$Astronomy Centre, University of Sussex, Falmer, Brighton BN1 9QH, UK\\
$^{11}$Institute for Astronomy, University of Edinburgh, Royal Observatory, Blackford Hill, Edinburgh, EH9 3HJ, UK\\
$^{12}$Jeremiah Horrocks Institute, University of Central Lancashire, Preston PR1 2HE, UK\\
$^{13}$Max Planck Institute for Nuclear Physics (MPIK), Saupfercheckweg 1, 69117, Heidelberg, Germany\\
$^{14}$School of Physics and Astronomy, Cardiff University, The Parade, Cardiff, CF24 3AA, UK\\
$^{15}$Sterrenkundig Observatorium, Universiteit Gent, Krijgslaan 281 S9, B-9000 Gent, Belgium\\
$^{16}$Universita' di Padova, Vicolo dell'Osservatorio, 3-35122, Italy\\
$^{17}$Instituto de Astrof´\u0131sica de Canarias, C/V´\u0131a L´actea s/n, E-38200 La Laguna, Spain\\
$^{18}$Department of Physics, Swiss Federal Institute of Technology (ETH-Z{\" u}rich), 8093 Z{\" u}rich, Switzerland\\
$^{19}$University of California, Irvine, Department of Physics \& Astronomy, 4186 Frederick Reines Hall, Irvine, CA 92697-4575\\
$^{20}$School of Astronomy, Institute for Research in Fundamental Sciences (IPM), PO Box 19395-5746, Tehran, Iran
$^{21}$Institute for Computational Cosmology, Department of Physics, Durham University, South Road, Durham, DH1 3LE, UK\\
$^{22}$Sterrenkundig Observatorium, Universiteit Gent, Krijgslaan 281 S9, B-9000 Gent, Belgium\\
$^{23}$Department of Physics and Astronomy, The Open University, Walton Hall, MK7 6AA Milton Keynes, UK\\
$^{24}$UK Astronomy Technology Centre, Royal Observatory Edinburgh, Edinburgh, EH9 3HJ, UK\\
$^{25}$Center for Astrophysics, Science \& Technology Research Institute, University of Hertfordshire, Hatfield, Herts\\
$^{26}$Leiden University, P.O.~Box 9500, 2300 RA Leiden, The Netherlands\\
$^{27}$Carnegie Institution for Science, 813, Santa Barbara Street, Pasadena, California, 91101\\
$^{28}$Institute of Cosmology and Gravitation (ICG), University of Portsmouth, Dennis Sciama Building, Burnaby Road, Portsmouth PO1 3FX, UK\\
$^{29}$School of Physics, Monash University, Clayton, Victoria 3800, Australia\\
$^{30}$Astrophysics Branch, NASA Ames Research Center, Mail Stop 245-6, Moffett Field, CA 94035, USA\\
$^{31}$Astronomy Unit, Queen Mary University London, Mile End Rd, London E1 4NS, UK\\
$^{32}$Leiden Observatory, NL-2300 RA Leiden, The Netherlands}
\begin{document}

\date{Accepted 2011 March 02}

\pagerange{\pageref{firstpage}--\pageref{lastpage}} \pubyear{2010}

\maketitle

\label{firstpage}
\clearpage
\begin{abstract}
We use multiwavelength data from the Galaxy And Mass Assembly (GAMA) and Herschel ATLAS
(H-ATLAS) surveys to compare the relationship between various dust obscuration measures in
galaxies. We explore the connections between the ultraviolet (UV) spectral slope, $\beta$,
the Balmer decrement, and the far infrared (IR) to $150\,$nm far ultraviolet (FUV)
luminosity ratio. We explore trends with galaxy mass, star formation rate (SFR) and redshift
in order to identify possible systematics in these various measures.
We reiterate the finding of other authors that there is a large scatter between the
Balmer decrement and the $\beta$ parameter, and that $\beta$ may be poorly constrained
when derived from only two broad passbands in the UV. We also emphasise that FUV
derived SFRs, corrected for dust obscuration using $\beta$, will be overestimated unless
a modified relation between $\beta$ and the attenuation factor is used. Even in the optimum
case, the resulting SFRs have a significant scatter, well over an order of magnitude.
While there is a stronger correlation between the IR to FUV luminosity ratio
and $\beta$ parameter than
with the Balmer decrement, neither of these correlations are particularly tight, and
dust corrections based on $\beta$ for high redshift galaxy SFRs must be treated with caution.
We conclude with a description of the extent to which the different obscuration
measures are consistent with each other as well as the effects of including other galactic properties on these correlations.
\end{abstract}

\maketitle

\section{Introduction}
Dust obscuration in galaxies is a well-recognised and long studied phenomenon. From the
perspective of large galaxy surveys, the primary concern is often how to make suitable corrections
for the dust obscuration within the galaxies of interest, in order to establish their intrinsic properties such as SFR and stellar mass.
This has traditionally been accomplished through the use of the Balmer decrement \citep{Ost:89},
a dust sensitive emission line ratio that is straightforward to measure in optical spectra. With the
advent of ultraviolet (UV) and infrared (IR) satellite telescopes, new approaches to measuring
or constraining global galaxy dust properties have been established, with common metrics being
the UV spectral slope, $\beta$ \citep{Meu:99}, and the far infrared (FIR) to ultraviolet luminosity ratio,
$L_{\rm FIR}/L_{\rm UV}$ \citep{Bell:03}.

The UV spectral slope $\beta$ has been proposed as a suitable tool for deriving obscuration
corrections, in particular for galaxies at high redshift where the Balmer decrement is not easily
measurable \citep[e.g.,][]{Bou:09}, having been shifted to infrared wavelengths. The effectiveness
of $\beta$ as an obscuration metric depends on how well the UV slope can be measured
\citep{Cal:94,Kong:04}. This is influenced by factors such as the instrumentation used and
associated sampling of the observed UV spectrum, together with the source redshift
impacting on the rest-frame UV wavelengths being probed.
From recent comparisons of $\beta$ with other obscuration measures, using samples
selected at optical \citep{Wij:10} or FIR \citep{Buat:10} wavelengths,
it is apparent that there are significant limitations in the use of the $\beta$ parameter
for making obscuration corrections.

The $\beta$ parameter was formalised by \citet{Meu:99} by exploiting the relationship between
the ratio of the far-infrared (FIR) and the ultraviolet (UV) fluxes and the UV spectral
slope for a sample of 57 starburst galaxies. \citet{Meu:99} argue that since the FIR flux in starburst galaxies is
produced from the UV radiation that is absorbed and re-radiated by dust, the FIR to UV flux ratio
can be used as a measurement of dust absorption. 
Subsequent work has emphasised that this relationship, originally calibrated for starburst galaxies, does not work so well 
for the general population of star forming galaxies \citep[e.g.,][]{Bell:02,Kong:04,Buat:05},
with typical galaxies deviating from the relationship between $\beta$ and
$L_{\rm FIR}/L_{\rm UV}$ established by \citet{Meu:99}. 

This is to be expected, as in quiescent spiral (non-starburst) galaxies both the assumption that the dust emission is solely powered by UV photons and the
foreground screen assumption breaks down. Thus as much as 30\% of the dust emission in spiral galaxies is predicted to come from the old stellar
populations \citep{Pop:11}. Furthermore, as shown in \citet{Pop:11}, at least 10\% of the dust heating could come from the old stellar population
in the bulge, for a typical  spiral with a bulge-to-disk ratio of around 0.3. Radiative transfer modelling of the distribution of stars and dust has also
shown that the finite exponential disk distribution of stars and dust, as opposed to the foreground screen distribution, could lead to significant
differences in the determination of the attenuation of stellar light \citep{Tuf:04,Pop:04}. Not least, spiral galaxies have
both a large scale distribution of stars and dust as well as a clumpy component associated with the star forming complexes in the spiral arms
\citep{Sau:05,Pop:10}. This renders a proper calibration of the $\beta$ relation to the radiative transfer models challenging.
Neverthless more empirical approaches have been taken, in the hope to find further parameters for calibrating this relation.

\citet{Kong:04}, therefore, point out that
the relationship between the IR-UV ratio and the UV slope is dependent on the star formation history as well as the dust content, implying that a straightforward
application of this relation might not be accurate. They present a modified relation which
accounts for the star formation history as well as the dust content. \citet{Wij:10} and \citet{Buat:10}
both find that SFRs derived from the FUV corrected for obscuration using $\beta$ and the
attenuation relation of \citet{Meu:99} are significantly overestimated. \citet{Wij:10}
propose a revised attenuation relation for $\beta$ which eliminates this overestimation.
Significant scatter, more than an order of magnitude, however, remains between the SFRs inferred
using dust corrections based on $\beta$ compared to those based on the Balmer decrement.

While the use of $\beta$ is attractive as a measure of dust absorption at high redshift,
in practice there are several considerations that need to be taken into account.
The main assumption made by \citet{Meu:99} was that high redshift galaxies will have similar
multiwavelength properties to local starbursts. It is unclear how the intrinsic spectral slope will be affected with increasing redshift, and indeed recent evidence for very blue UV spectral slopes
at the highest redshifts \citep{Bou:09,Bou:10,Bun:10} is now being interpreted as evidence potentially
for either extremely low metallicity or variations in stellar initial mass function. This opens the
question of how much intrinsic variation there is in the underlying UV spectral slope amongst
galaxies at progressively lower redshift, in particular if recent suggestions for variation of
the stellar initial mass function \citep{Dav:08,Wil:08a,Wil:08b,Meu:09,Gun:11} are borne out.
If there is indeed broad variation in the UV spectral slope, then there needs to be a redshift, or potentially a SFR or other dependence included in the use of $\beta$ when making
obscuration corrections. 


To address these issues, we explore the relationships between the FIR luminosity,
the UV luminosity, the Balmer decrement and $\beta$, all as a function
of galaxy mass, SFR, and redshift. In \S\,\ref{data} we describe the data used in this analysis.
\S\,\ref{BD} presents an analysis of the $\beta$ and Balmer decrement relationships, and
\S\,\ref{uvatt} explores different approaches to parameterising the UV attenuation.
\S\,\ref{FIRFUV} investigates the relationships between the FIR and FUV luminosities. These
measures are discussed and analysed in \S\,\ref{disc} and in \S\,\ref{conc} we
summarise our results and conclusions. Throughout, all magnitudes are given in the AB system,
and we assume a cosmology with
$H_{0}=$70\,km\,s$^{-1}$\,Mpc$^{-1}$, $\Omega_{M}=0.3$ and $\Omega_{\Lambda}=0.7$.

\section{Data}
\label{data}
\renewcommand{\thefootnote}{\fnsymbol{footnote}}
The Galaxy And Mass Assembly (GAMA)\footnote[2]{http://www.gama-survey.org/}
survey is a multiwavelength imaging and spectroscopic
survey covering $\approx 144$ square degrees of sky in three $12^{\circ} \times 4^{\circ}$
regions \citep{Dri:09,Dri:10,Rob:10,Bal:10,Hill:10}. GAMA provides the redshifts, emission line
measurements and UV/optical/near IR (NIR) photometry used in this analysis, with spectroscopy from
2dF/AAOmega on the Anglo-Australian Telescope and imaging from GALEX, SDSS
and UKIDSS. The FUV band in the GALEX filter has an effective wavelength of $1528\,$\AA\ and
the NUV band has an effective wavelength of $2271\,$\AA. $K$-corrections were applied to the
observed GALEX UV magnitudes using {\sc kcorrect.v4.1.4} \citep{Bla:03} to infer the rest-frame magnitude at $1528\,$\AA. The galaxies in this analysis have a GALEX FUV magnitude range of
$17.2 < m_{FUV} < 26.8$, and a redshift limit of $z < 0.35$ due to the requirement
for H$\alpha$ to be in the observable spectral range. Stellar masses are calculated for the
GAMA galaxies using spectral energy distribution (SED) models based on the $ugriz$ photometry bands \citep{Tay:10}.

The \emph{Herschel}\footnote[3]{Herschel is an ESA space observatory with science instruments provided
by European-led Principal Investigator consortia and with important
participation from NASA}-$\mathrm{ATLAS^2}$ (H-ATLAS)\footnote[4]{http://www.h-atlas.org/} survey will ultimately observe 550 square degrees of sky in 5 bands (100, 160, 250, 350, $500\,\mu$m) 
detecting $\sim 200,000$ 
galaxies spanning $0<z<4$ (Eales et al. 2010). The H-ATLAS Science Demonstration Phase (SDP) field covers an area of $\sim 4^{\circ} \times 4^{\circ}$ centered at 
$\alpha=09^h 05^m,\, \delta=00^{\circ} 30^{\prime}$ in the GAMA 9hr field. It was mapped in parallel mode using the PACS \citep{Pog:10} and SPIRE 
(Griffin et al. 2010) instruments on-board the {\em Herschel Space Observatory\/} \citep{Pil:10}. The 5-band {\em Herschel\/} maps and catalogues 
for this SDP field are described in \citet{Iba:10}, \citet{Pas:10} and \citet{Rig:10}. Sources in the $>5 \sigma$ $250\,\mu$m catalogue were 
matched to the SDSS DR7 (Abazajian et al. 2010) and thence to GAMA using a likelihood ratio method \citet{Smi:10} resulting in 2423 250$\,\mu$m sources 
with optical identifications from SDSS to $r<22.4$ at $>80$ percent reliability. Of these sources, 1050 were found to be in common with the GAMA survey in 
the $\sim 12.7$ square degrees of overlap. From the reliability values calculated for these objects, we expect to have a 1.9 percent false ID rate in this sample.

\begin{figure}
\centerline{\includegraphics[width=90mm]{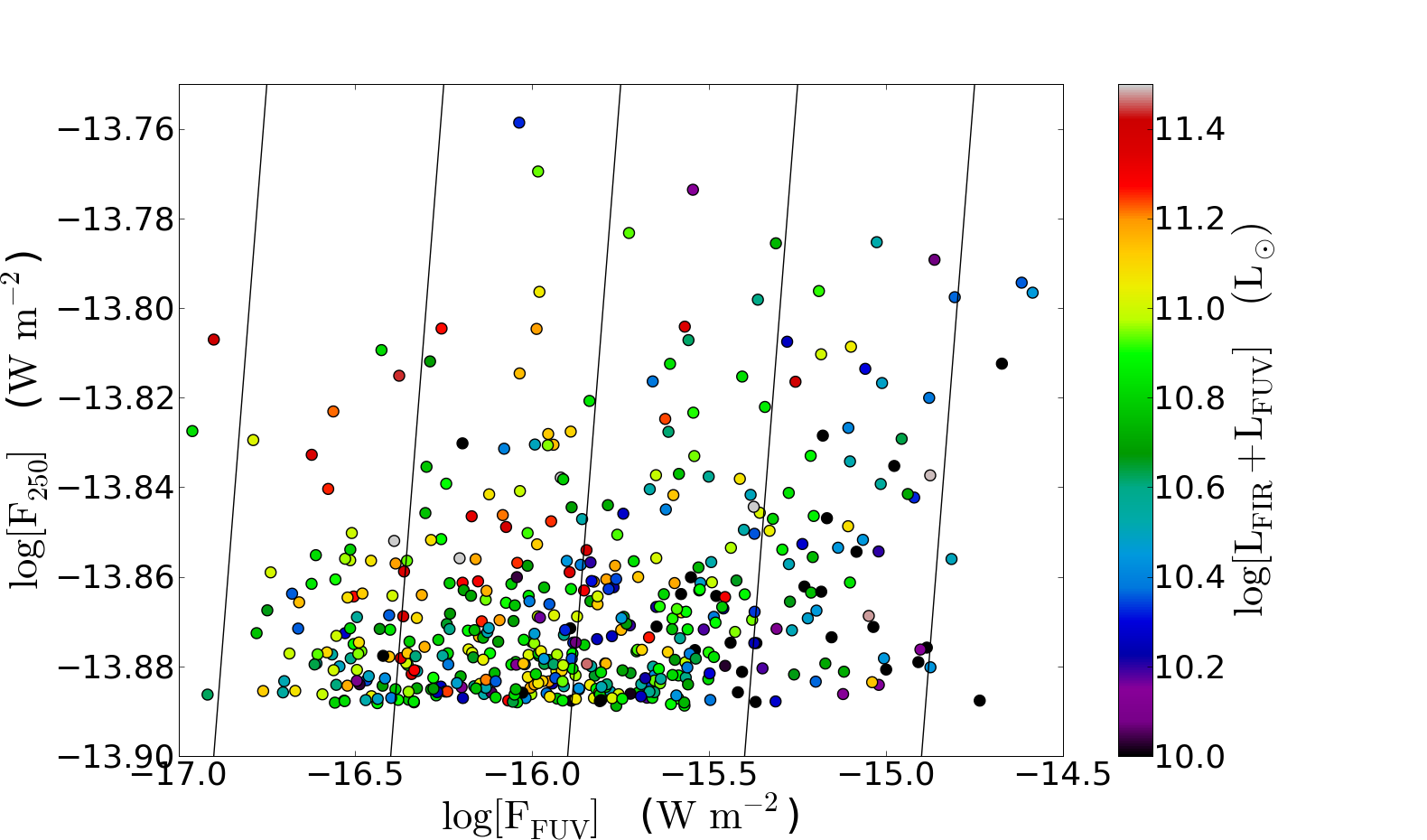}}
\caption{250$\mu$m flux compared against the FUV flux, illustrating the flux limits of the GALEX
and H-ATLAS measurements in the current analysis. Both quantities are in units of
W\,m$^{-2}$, and the lines are constant values of $\log(F_{\rm 250}/F_{\rm FUV})$.
The colour indicates the total of the IR and FUV luminosity.}
\label{fig:ir_vs_fuv_flux}
\end{figure}

The IR and FUV luminosities that appear independently have also followed the same treatment.
The energy balance method of da Cunha, Charlot \& Elbaz (2008) is used to derive
the IR (dust) luminosity from 3-1000$\mu$m by fitting to the
the UV-sub-mm spectral energy distributions of the galaxies.
A large, stochastic library of stellar and dust emission 
models was used that includes a wide range of star formation histories, metallicities, dust attenuation, dust 
temperatures and different contributions by various dust emission components to the total infrared emission
\citep[polycyclic aromatic hydrocarbons (PAHs), hot mid-infrared continuum, and dust in thermal equilibrium; full details are given in
in][]{Cun:08}. Throughout, we use the notation $L_{IR}$ to represent the IR luminosity described above.
Figure~\ref{fig:ir_vs_fuv_flux} shows the UV and $250\,\mu$m fluxes,
colour-coded by the sum of the FUV and IR luminosities, along with lines of constant
flux ratio, to illustrate the range of values that this sample probes.
The sum of the FUV and IR luminosities are derived from frequency independent luminosities of these two quantities (units of 
solar luminosities). 

The data used in the SED fitting includes UV photometry from GALEX (Seibert et al. in
prep.), optical {\em ugriz\/} from SDSS DR7 (Abazajian et al. 2009),
infrared {\em YJHK\/} from UKIDSS LAS (Lawrence et al. 2007) all of which is re-measured from the images in matched apertures defined in the r-band by GAMA (Driver et al. 2010; Hill et al. 2010).
The optical/NIR fluxes have been PSF (point spread function) matched \citep{Hill:10}, and the UV fluxes were derived using the the technique of \citet{Rob:11}.
See Smith et al. (in prep.) for a full description of the implementation of this method to H-ATLAS data.

Of the 1050 GAMA sources detected within the H-ATLAS SDP region, 899 have H$\alpha$
emission, and 875 have FUV measurements from the GALEX-MIS survey.
Of these galaxies 221 have measurements from both the PACS ($\sigma > 5$) and SPIRE instruments on the Herschel telescope
while the rest have only measurements from the SPIRE instrument. 
Within the GAMA survey there are 96231 galaxies with H$\alpha$ measurements and
104681 with FUV measurements. Where we show measurements that do not rely
on having IR data, we use the full GAMA sample for which H$\alpha$ emission and UV photometry have been
measured, after excluding AGN based on spectral emission line ratio diagnostics \citep{Bld:04, Kwl:01}. Where the IR luminosities are required, we are limited to showing only those
within the H-ATLAS SDP region.

\section{Balmer decrement and UV spectral slope}
\label{BD}
The Balmer decrement is the ratio of the stellar absorption corrected H$\alpha$ and H$\beta$ fluxes.
For the GAMA emission line measurements, the stellar absorption corrections are applied as:
\begin{equation}
f_s = \left[\frac{EW + EW_{c}}{EW}\right]f_{o},
\end{equation}
where $f_{o}$ and $f_{s}$ are the observed and stellar absorption corrected fluxes respectively,
EW is the equivalent width of the emission line and $EW_{c}$ is the
correction for stellar absorption, taken to be 0.7\,\AA\, \citep{Gun:11}.
\citet{Gun:11} show that varying $EW_{c}$ for each galaxy will not have a 
significant effect on the observed trends. They show that a 5\% difference in H$\alpha$ EWs are observed, and only for the extremely low
(log[H$\alpha$ EW] $< 0.9$) H$\alpha$ EWs. For this same reason, varying $EW_{c}$ values for different emission lines will also have a minimal effect.
We, therefore opt to use a standard $EW_{c}$ value of 0.7\AA .
The Balmer decrement is then
\begin{equation}
BD = f_{H\alpha} / f_{H\beta},
\end{equation}
where BD is the Balmer decrement and $f_{H\alpha}$ and $f_{H\beta}$ are the stellar
absorption corrected fluxes of the H$\alpha$ and H$\beta$ emission lines.
Aperture corrections were performed on emission line data following the
prescription of Hopkins et al. (2003; see also Gunawardhana et al. 2011).

All Balmer decrements below the Case-B value of 2.86 \citep{Ost:89} were set equal to 2.86 as suggested by \citet{Kwl:06}.
The application of the Balmer decrement in correcting the H$\alpha$ luminosities
for dust obscuration is described in \citet{Wij:10}, and we use the same method here. In
particular we adopt the obscuration curve of \citet{FD:05}, shown by \citet{Wij:10} to
be the most effective at producing self-consistent SFR estimates simultaneously from
FUV, NUV, [OII] and H$\alpha$ luminosities.

The $\beta$ parameter is calculated using the observed fluxes obtained through the GALEX
FUV and NUV filters. The UV spectral slope is determined from a power-law fit to the UV
continuum of the form: 
\begin{equation}
f_{\lambda} \propto \lambda^{\beta},
\end{equation}
where $f_{\lambda}$ is the flux density per wavelength interval and $\lambda$ is the central rest wavelength \citep{Meu:99}. Following \citet{Kong:04} we use:
\begin{equation}
\beta=\frac{log{{\bar{f}}_{FUV}} - log{{\bar{f}}_{NUV}}}{{log{\lambda}}_{FUV} - {log{\lambda}}_{NUV}},
\end{equation}
where ${\lambda}_{FUV} = 1528\,$\AA\ and ${\lambda}_{NUV} = 2271\,$\AA\ are the effective
wavelengths of the far and near UV filters of the GALEX satellite and ${\bar{f}}_{FUV}$ and
${\bar{f}}_{NUV}$ are the mean flux densities per unit wavelength through these filters.

\begin{figure}
\centerline{\includegraphics[width=62mm, height=60mm]{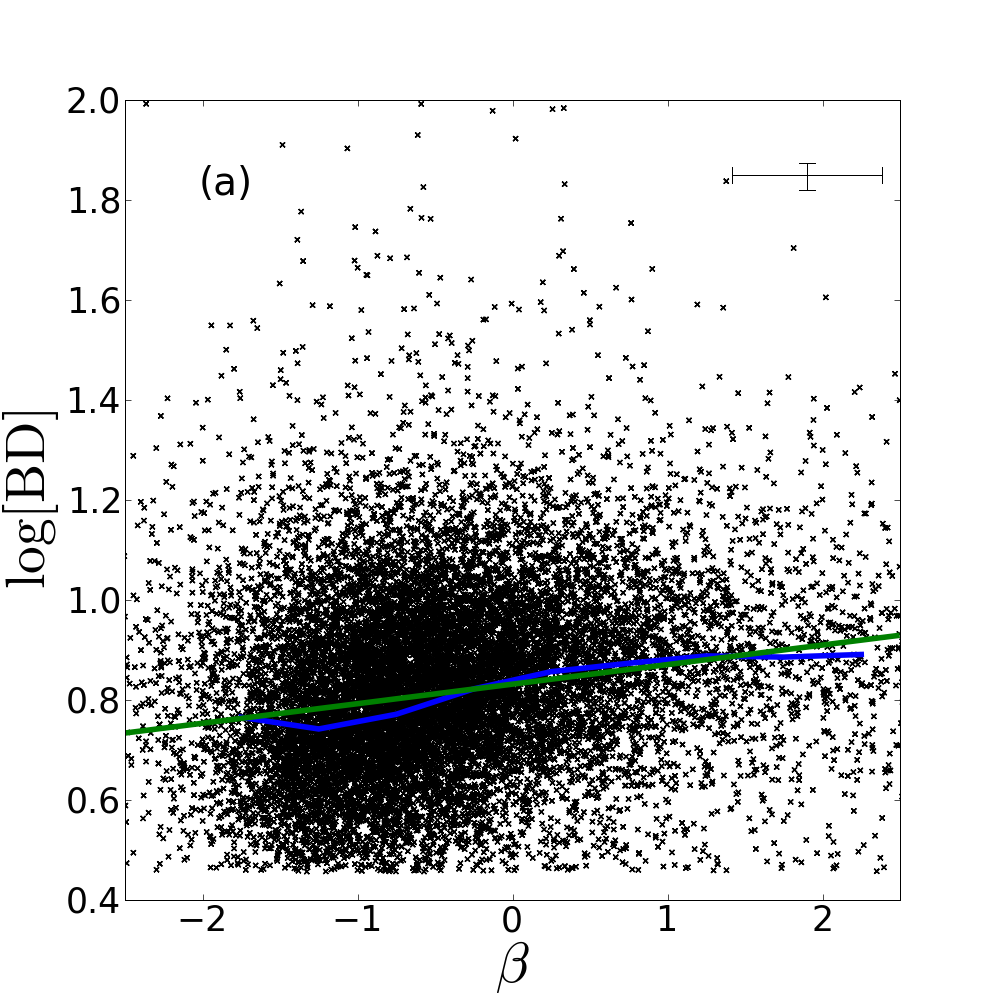}}
\centerline{\includegraphics[width=62mm, height=60mm]{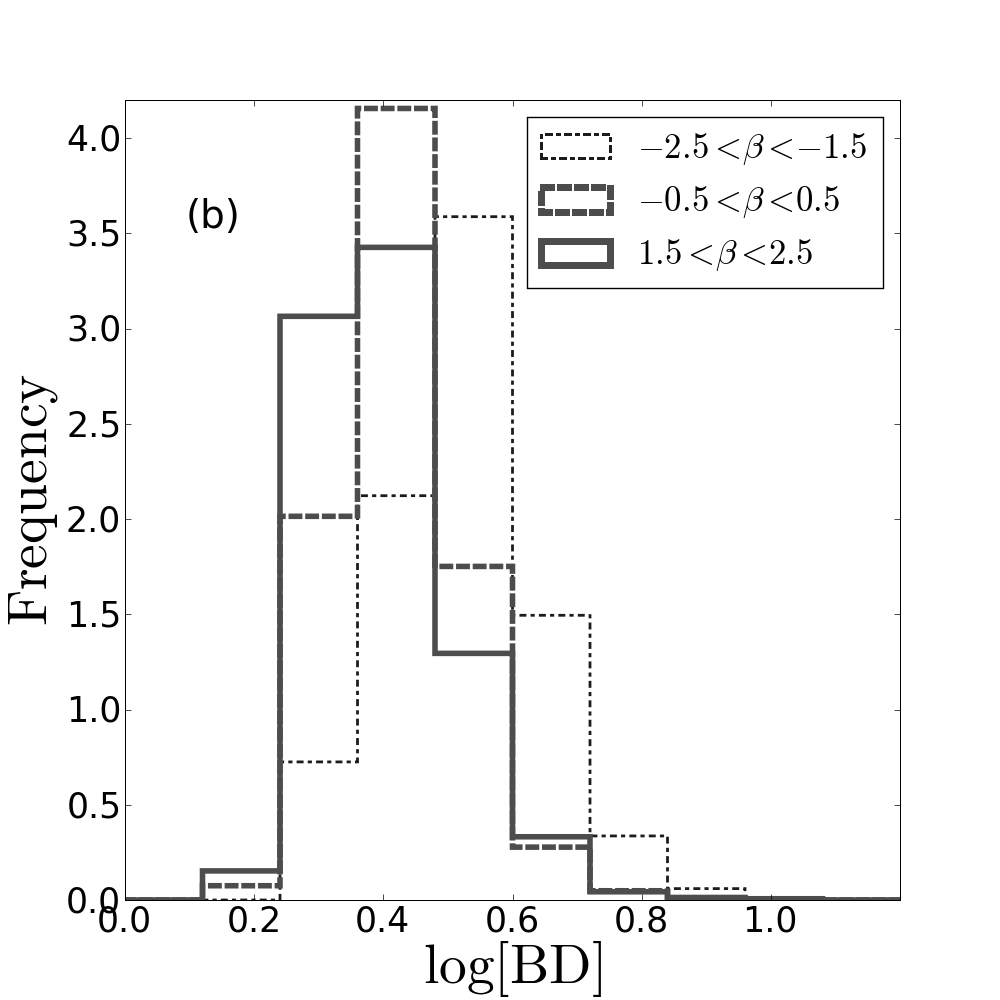}}
\caption{(a) Comparison of Balmer decrement and the $\beta$ parameter. There is a very
weak trend, although for any given value of $\beta$ the Balmer decrement can vary
by factors of typically 2-3, but ranging up to ten. The blue line shows the mean values for the Balmer decrement in bins of $\beta$
while the green line shows the line of best fit. The error bars show the median errors.
(b) These histograms show how the Balmer decrements are distributed across three $\beta$
ranges. As we move to higher $\beta$ values it is clear that the typical Balmer decrement also increases. The 3 bins contain 2510, 9788 and 1115 galaxies respectively
in the order of lowest to highest $\beta$ ranges. Two bins ($-1.5<\beta<-0.5$ and $0.5<\beta<1.5$) are omitted for clarity, although they follow the same trend.}
\label{bds}
\end{figure}

Figure~\ref{bds} shows a very weak trend between the Balmer decrement and $\beta$, but this is
dominated by the remarkably large scatter in the distribution. The least-squares fit shown,
with an unremarkable correlation coefficient of $r=0.18$, is $BD=0.64\beta+7.64$. While
we do not claim any formal correlation between these two measures, we use this result below
in an exploration of different parameterisations of the FUV attenuation, $A_{FUV}$. In order to minimise any bias in the least squares
fit any Balmer decrements with values $\leq 2.86$ were excluded from the fit. This removes $\approx$ 21\% of the sample for the purposes of this fit.

While there is a large scatter, it can be seen from Figure~\ref{bds} that at the highest $\beta$ values there is a concentration of
systems with higher Balmer decrements. In other words, systems with low Balmer decrements tend to have flatter UV spectral slopes.
It is also interesting to note that the lowest values of $\beta$ (the bluest UV spectral slopes) are not associated with the lowest
Balmer decrements. Galaxies with the highest measured Balmer decrements tend to have a broad range of UV spectral slopes. 
Systems with flat UV slopes ($\beta\sim 0$), however, have the broadest range of Balmer decrement.

The Balmer decrement and the $\beta$ parameter both measure the strength of obscuration in
different ways, so naively a correlation between the two parameters might be expected.
The dust geometry in galaxies is not a simple foreground screen, though, instead having
complex filamentary and patchy structure \citep{Cal:97,Cal:01}. A consequence of this
is that Balmer decrement and $\beta$ are likely to be sensitive to the average obscuration
at different optical depths within a galaxy. Indeed even H$\alpha$ and H$\beta$ can be seen to probe different optical depths \citep{Sjn:02}.
It is also the case that the UV spectral slope
includes emission from older stellar populations than the OB stars responsible for ionising the
hydrogen \citep[e.g.,][]{CF:00, Bell:03}, and that these older stellar populations are likely to be distributed
differently throughout a galaxy than the OB stars, and hence being affected by different
levels of dust obscuration \citep[e.g.][]{Cal:97}. In the absence of sophisticated modelling of the
radiative transfer within galaxies \citep[e.g.,][]{Wit:92,Xil:97,Sil:98,Pop:00,Tuf:04,Pop:11} it is challenging to overcome this
limitation.

In the following sections we analyse the distribution of galaxy SFR, mass and redshift, on the relationship between the 
Balmer decrement, the $\beta$ parameter and the IR and FUV luminosities.

\section{UV attenuation}
\label{uvatt}
We use $L_{IR}/L_{FUV}$ as a measure of the UV attenuation under the assumption that the energy absorbed from the UV luminosity is re-emitted in the IR.
We would naively expect to see higher BDs and steeper UV slopes with increasing $L_{IR}/L_{FUV}$.

\begin{figure}
\centerline{\includegraphics[width=90mm, height=90mm]{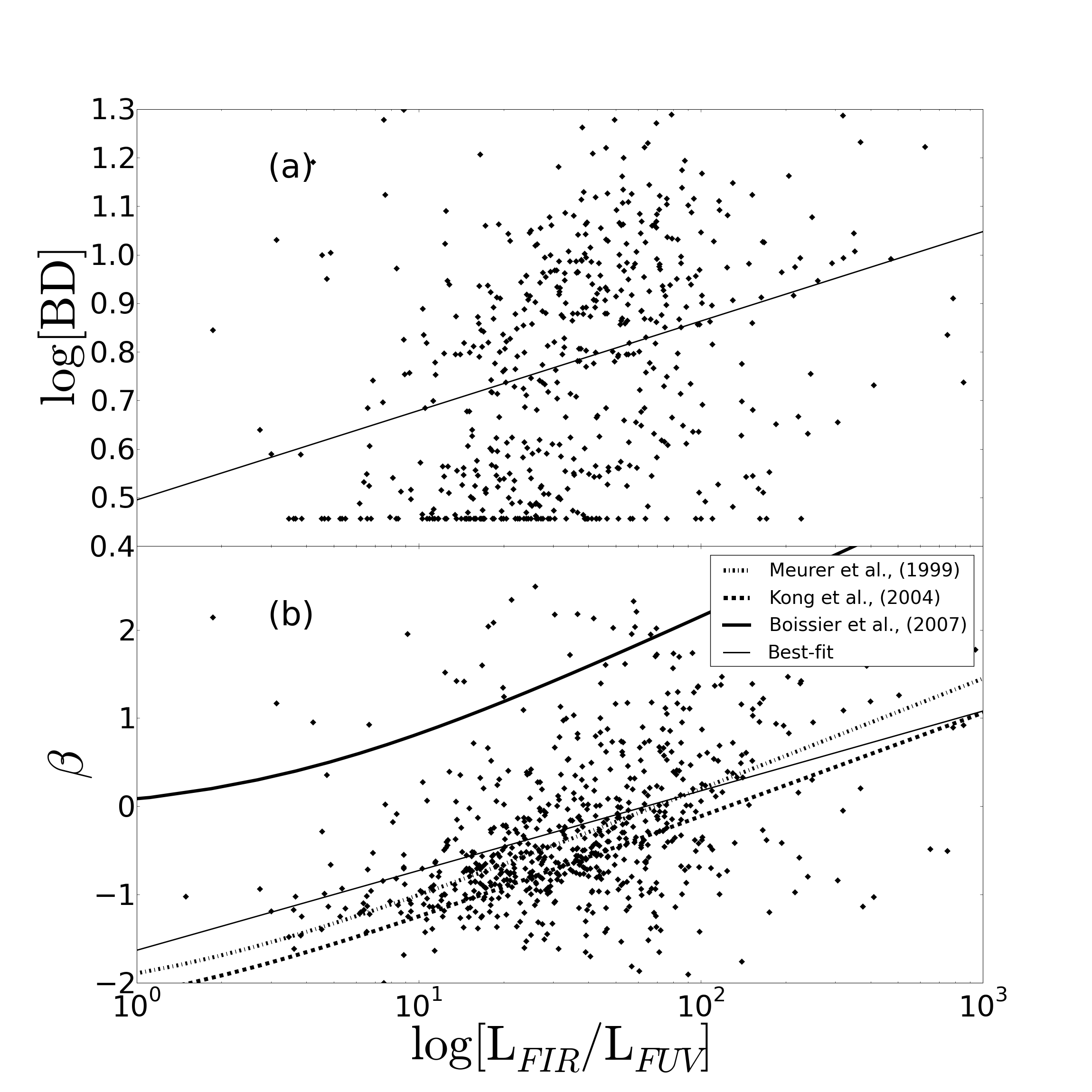}}
\caption{(a) Balmer decrement and (b) $\beta$, as a function of the
IR to dust uncorrected FUV luminosity ratio. The thin solid lines in each panel show the
least-squares linear best fit. Other relationships between $\beta$ and $L_{IR}/L_{FUV}$
are shown as follows. Dashed line: \citet{Kong:04} for local starbursts; Dash-dot line: \citet{Meu:99} also for local starbursts;
Thick solid line: \citet{Boi:07} for optically selected star forming galaxies.}
\label{bds2}
\end{figure}

Figure~\ref{bds2} shows the variation of both Balmer decrement and $\beta$ as a
function of $L_{IR}/L_{FUV}$. These relationships show weak trends, emphasised by
the best fit lines, although it is clear that there is a large scatter. The correlation coefficients
demonstrate that these trends are weak at best, with $r=0.44$ for Balmer decrement, and
$r=0.48$ for $\beta$, against $L_{IR}/L_{FUV}$.
It is encouraging to see that for low $L_{IR}/L_{FUV}$
the values of both Balmer decrement and $\beta$ are small, although at
high $L_{IR}/L_{FUV}$ the range of possible obscuration values becomes quite large
(although with a clearly increasing mean value as $L_{IR}/L_{FUV}$ increases).
Figure~\ref{bds2}(b) also shows the relationships found for local starbursts \citep{Meu:99,Kong:04} and
optically selected star-forming galaxies \citep{Boi:07}. Our sample, as also found for higher IR
luminosity systems by \citet{Buat:10}, spans the regime between these models. This is obviously
the origin of the discrepancies seen \citep{Wij:10,Buat:10} when applying the standard relations
between $\beta$ and attenuation in deriving SFRs.

\begin{figure}
\centerline{\includegraphics[width=70mm]{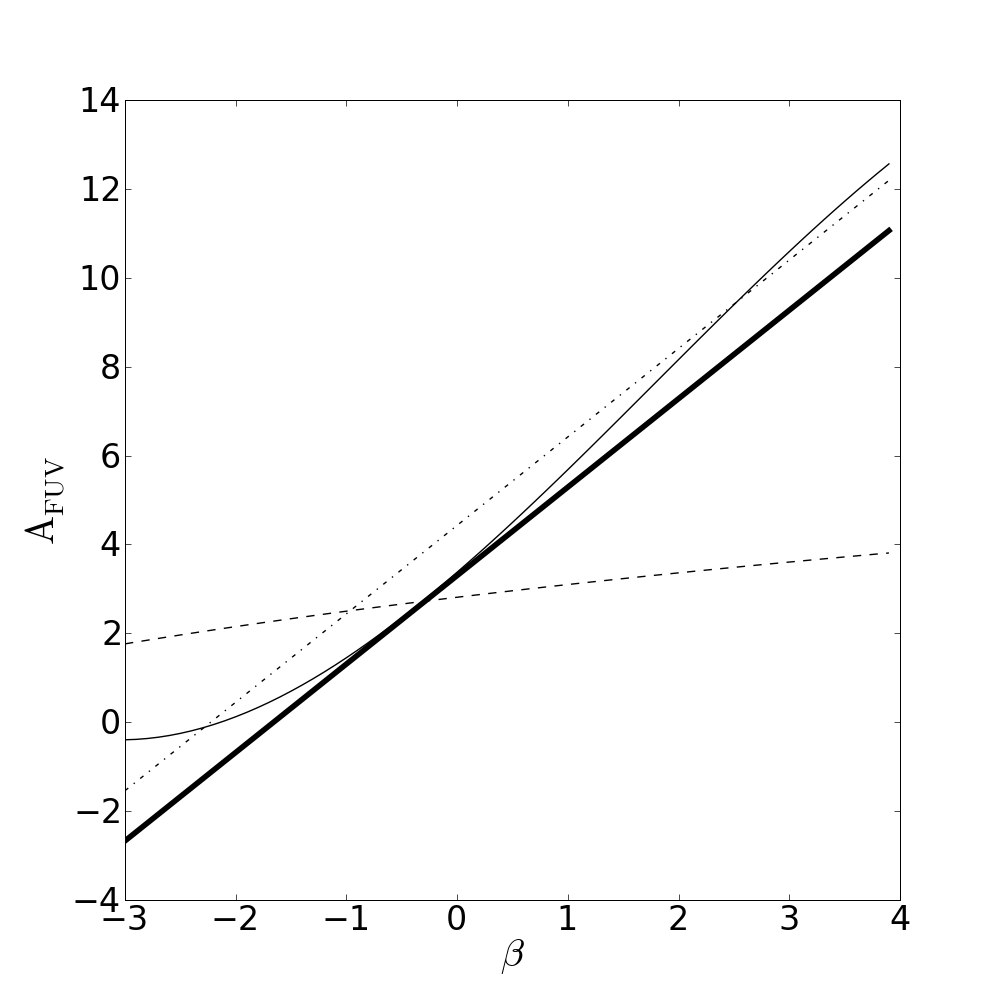}}
\caption{Different observational correlations between $\beta$ and the attenuation
at FUV wavelength, $A_{FUV}$. Thin solid line: Equation~2 from \citet{Buat:05} combined with Equation~\ref{ratiotobeta}; Dot-dash line: \citet{Meu:99};
Thick solid line: \citet{Wij:10}; Dashed line: This work, Equation~\ref{bdtoafuv} derived using the relationship between BD and $\beta$.}
\label{afuvvsbeta}
\end{figure}

It is instructional to look at how different approaches to parameterising the FUV attenuation,
$A_{FUV}$, as a function of $\beta$ compare. We show a number of different results
in Figure~\ref{afuvvsbeta}. In addition to the parameterisations of \citet{Meu:99} and
\citet{Wij:10}, we use our least-squares fit between $L_{IR}/L_{FUV}$ and $\beta$
from Figure~\ref{bds2}, 
\begin{equation}
\label{ratiotobeta}
log[L_{IR}/L_{FUV}]=(\beta + 1.66)/0.98,
\end{equation}
combined with Equation~2 from \citet{Buat:05}, to derive the thin solid line in Figure~\ref{afuvvsbeta}.
We also use the relation between Balmer decrement and $\beta$ derived in \S\,\ref{BD}
above, together with the relationship between Balmer decrement and $A_{FUV}$ arising
from the application of an obscuration curve \citep[here we use][]{FD:05}, to derive
\begin{equation}
\label{bdtoafuv}
A_{FUV}=7.49\log\left(\frac{0.62\beta + 6.79}{2.86}\right).
\end{equation}

\begin{figure}
\centerline{\includegraphics[width=70mm]{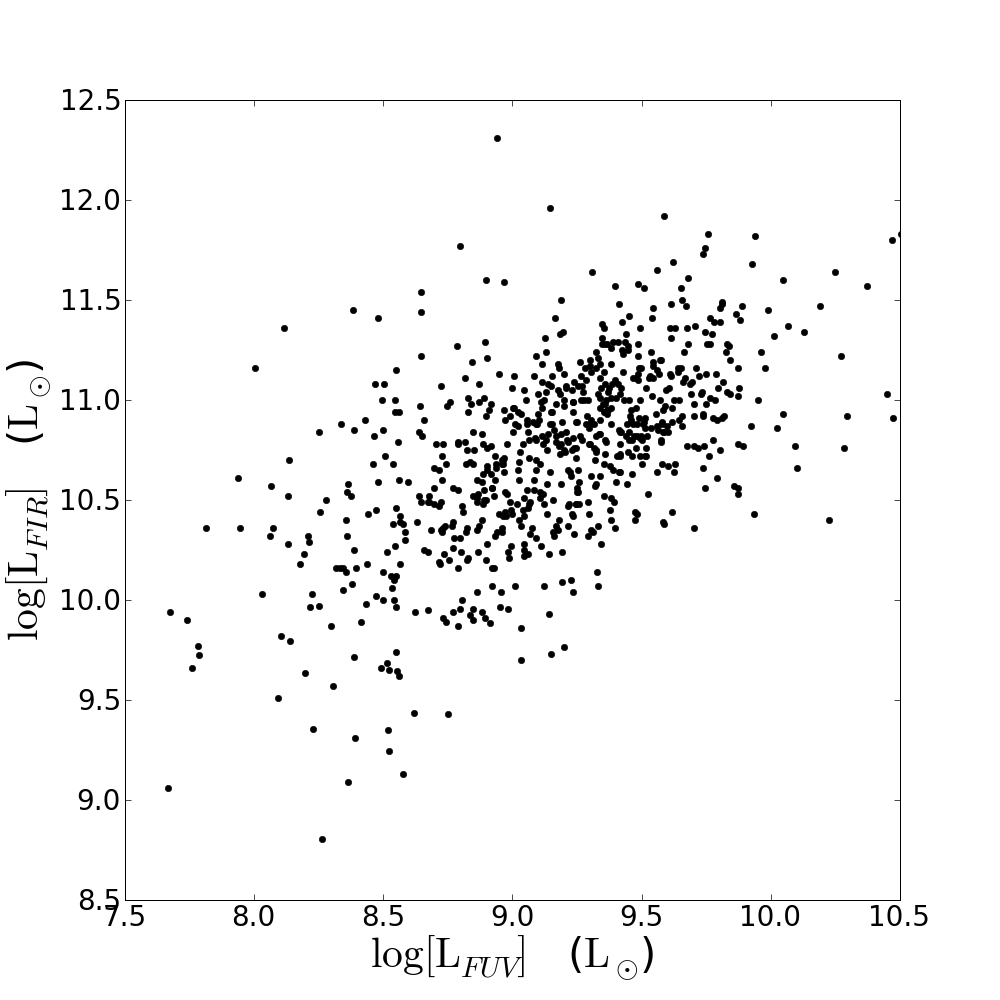}}
\caption{FUV luminosity compared with IR luminosity. While there is a clear relationship overall between the two luminosities, at low FUV luminosities there is a much broader range of
IR luminosities compared to high FUV luminosities.}
\label{lfuv_lfir}
\end{figure}
It is encouraging to see the consistency between the relationships of \citet{Buat:05} and
\citet{Wij:10}, and the offset of the relationship of \citet{Meu:99} has been discussed
above. It is particularly intriguing, however, to see the very different relationship presented
when Balmer decrement is used as a proxy in the process of estimating attenuation.
The significantly different slope derived here (dashed line in Figure~\ref{afuvvsbeta}), a consequence of the logarithmic dependence
on the Balmer decrement, may start to give some hints as to why such dramatically
different results are obtained when using these different approaches to obscuration
corrections in deriving SFRs. We return to this in \S\,\ref{sfrs} below.

\section{IR and FUV luminosities}
\label{FIRFUV}
Figure~\ref{lfuv_lfir} shows a clear trend between the FUV and IR luminosities.
The scatter within the trend varies at different FUV luminosities.
At high FUV luminosities the distribution of IR luminosities is relatively constrained ($\approx 1$ dex)
while at low FUV luminosities the distribution is much broader ($\approx 2.5$ dex).

\begin{figure}
\centerline{\includegraphics[height=180mm,width=80mm]{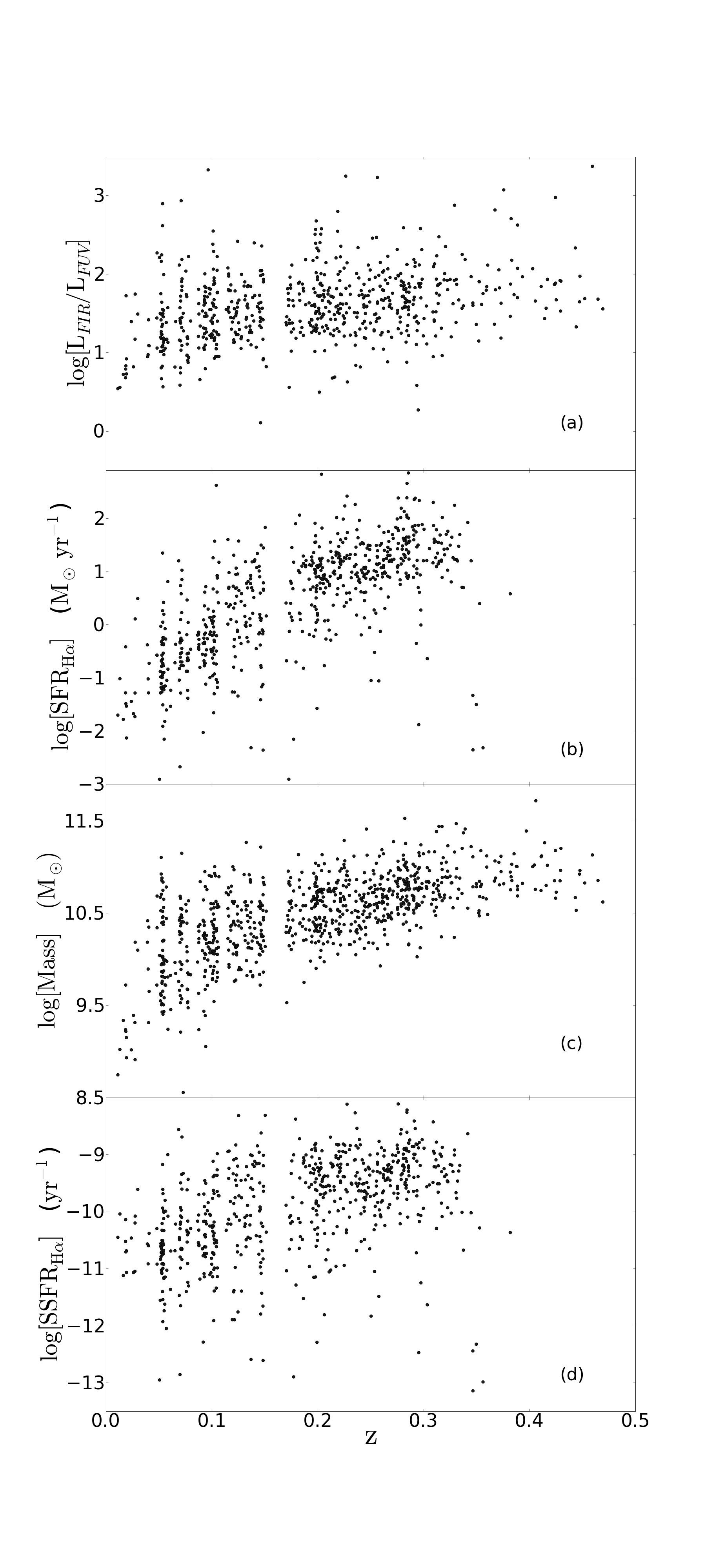}}
\caption{Variation with redshift of (a) $L_{IR}/L_{FUV}$ ratio, (b) H$\alpha$ SFR, (c) Mass, (d) specific H$\alpha$ derived SFR. The gap centred on z = 0.16
shows where the wavelength of the atmospheric $O_{2}$ band (Fraunhofer A-line) overlaps with redshifted wavelength of the
H$\alpha$ emission line. For this reason we cannot use H$\alpha$ emission line measurements that fall into this redshift range, leading us to omit these data from our analysis.}
\label{z_other}
\end{figure}

Figure~\ref{z_other} shows relationships between $L_{IR}/L_{FUV}$, H$\alpha$ SFR, mass and the specific SFR (SSFR) as a function of redshift, where SSFR is the SFR divided by the stellar mass of the
galaxy. The range of observable values for these properties becomes more limited with redshift, a consequence of the flux limited selection of the GAMA survey. This effect is strongest for SFR and 
mass, but less limiting for $L_{IR}/L_{FUV}$. The mild increase seen in average $L_{IR}/L_{FUV}$ with redshift is consistent with the evolution in dust mass found by Dunne et al. (2010).

In Figure~\ref{bigfig}, we compare the IR to FUV luminosity ratio, which indicates dust attenuation, against the
IR luminosity, the dust uncorrected FUV luminosity, and the sum of these two quantities,
in order to explore these relations as a function of SFR, mass and redshift. Each set of three panels
in Figure~\ref{bigfig} shows the same data, but colour-coded by different parameters in each
row to highlight the impact of these properties. Presenting the combination of $L_{FUV}$ and $L_{IR}$ data in this way
allows ease of comparison to the wealth of existing similar analyses \citep[e.g.,][]{Buat:05,Buat:09,Buat:10}.

It has become common now to infer SFRs by combining the IR and UV luminosities, as
an alternative to explicit obscuration corrections \citep[e.g.,][]{BX:96,Flo:99,Buat:99,Hop:06}.
The ranges of the values in Figure~\ref{lfuv_lfir} also agree with those observed in \citet{Buat:09}.
\citet{Buat:10} measure the IR to FUV luminosity
ratio for a sample of galaxies in the Lockman field surveyed by {\em Herschel\/} as part of
the HerMES\footnote[5]{http://www.hermes.sussex.ac.uk} survey. Our sample probes lower
infrared luminosities than \citet{Buat:10}, and also shows a slightly lower range in the
ratio of IR to FUV luminosities, a consequence of the low redshift range of our sample.

\begin{figure*}
\centerline{\includegraphics[width=200mm, height=200mm]{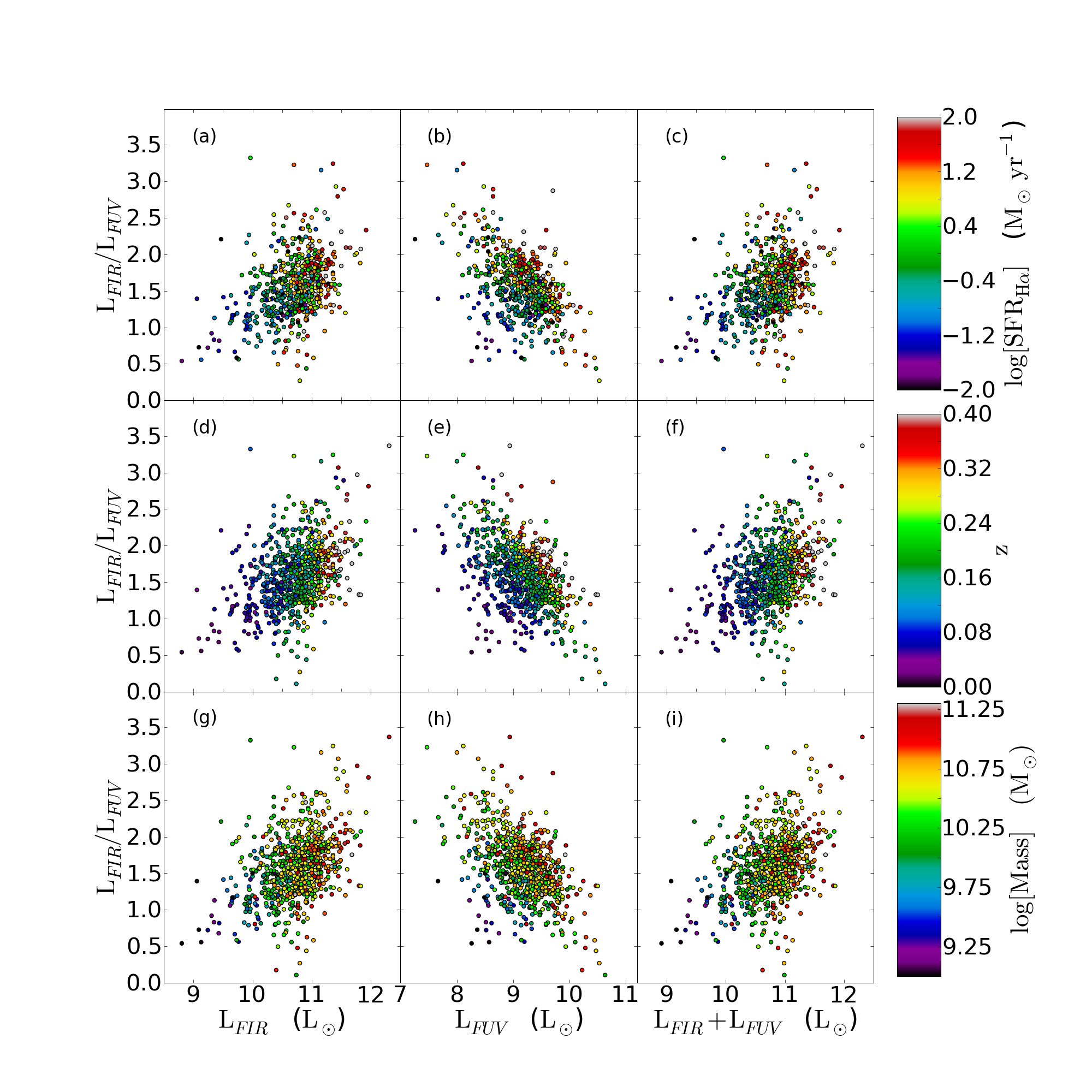}}
\caption{The IR to FUV luminosity ratio is shown against (from left to right),
the IR luminosity, dust uncorrected FUV luminosity, and the
sum of the IR and dust uncorrected FUV luminosity. The plots are colour coded as a
function of (from top to bottom) the dust corrected H$\alpha$ derived SFR, redshift, galaxy stellar mass, dust corrected H$\alpha$ derived SSFR, $\beta$ and Balmer decrement.
(Continued below.)}
\label{bigfig}
\end{figure*}

\addtocounter{figure}{-1}
\begin{figure*}
\centerline{\includegraphics[width=200mm, height=200mm]{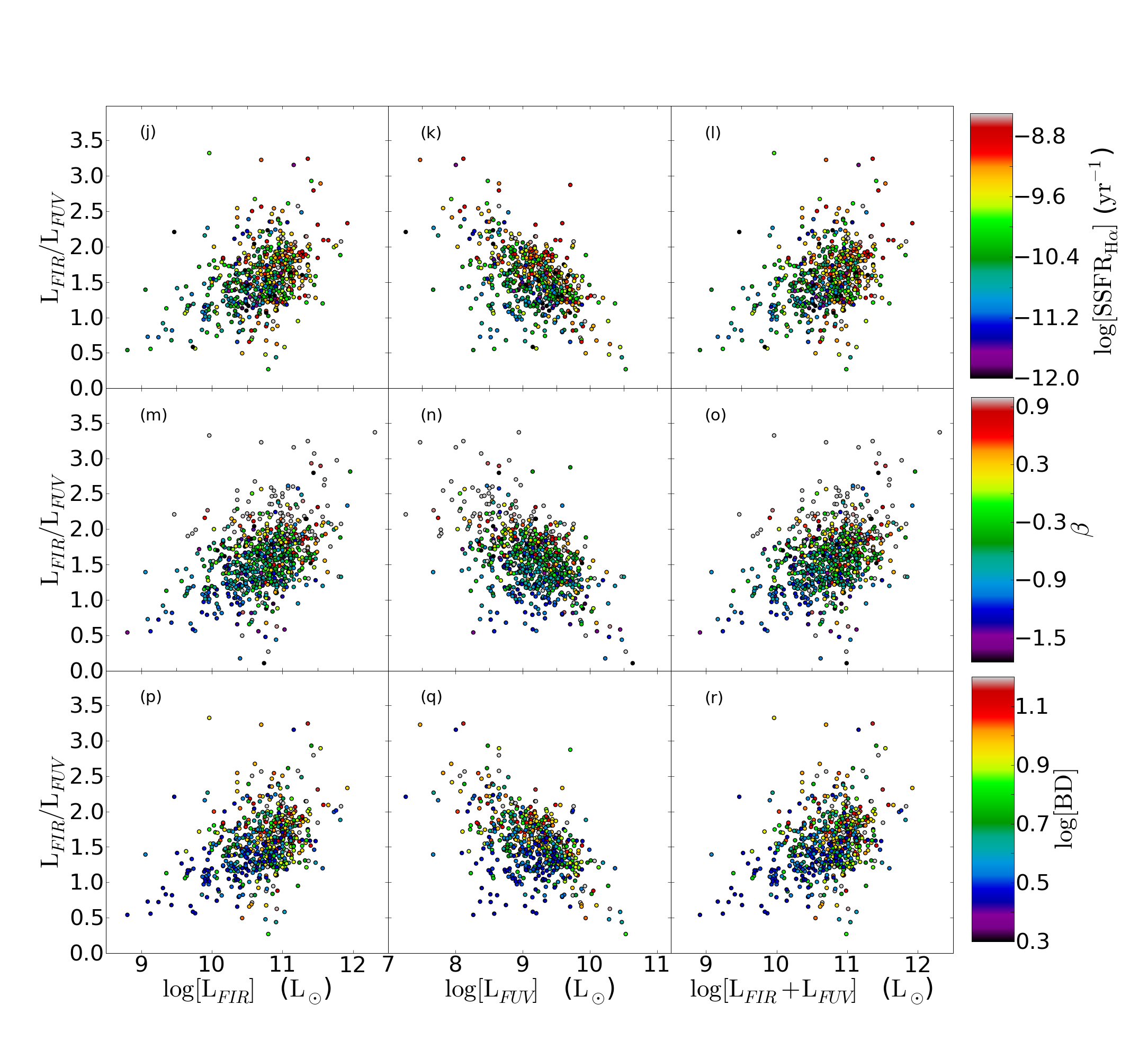}}
\caption{continued.}
\label{bigfig2}
\end{figure*}

The basic structure seen here is that systems
of low $L_{IR}$ have lower $L_{IR}/L_{FUV}$ ratios, while for high $L_{IR}$ a broad
range of $L_{IR}/L_{FUV}$ ratios is visible, with medians increasing with increasing $L_{IR}$
\citep[a trend which continues to higher $L_{IR}$, as in Figure~1 of ][]{Buat:10}.
It is clear that despite the broad range of ratios at high $L_{IR}$ the majority of galaxies display high $L_{IR}/L_{FUV}$ ratios.
Strikingly, the highest $L_{FUV}$ systems (prior to dust correction) are those with the smallest 
values of $L_{IR}/L_{FUV}$, a result that has been emphasised by \citet{Buat:09}, and which
highlights the limitations of UV-selection when performing a star-formation census.
Systems with low observed $L_{FUV}$ tend to have the highest $L_{IR}/L_{FUV}$ ratios,
an indication of significant obscuration in these systems. As a function of the combined
IR and FUV luminosity, the general trend is similar to that with the $L_{IR}$ alone. The
contribution of the FUV luminosity to the total is significant only for the lower $L_{IR}$ systems,
and essentially negligible for the more luminous infrared systems.

\citet{Buat:09} shows that the IR to UV luminosity ratio does not change significantly with increasing redshift for the majority of UV-selected galaxies. 
Interestingly, Lyman break galaxies at $z < 1$, which are photometrically colour selected, were found by \citet{Buat:09} to show
systematically lower $L_{IR}/L_{LUV}$ at the same UV luminosity compared to the UV-selected systems.
The implication here is that Lyman break galaxies maybe more biased against obscured systems than those
identified by simple UV selection. For IR selected (250$\mu$m selections with the SPIRE instrument on board the Herschel telescope) galaxies, however, Dunne et al. (2010) find that the dust mass
function evolves strongly with redshift, so that the dustiest galaxies at higher redshifts have higher characteristic dust masses than those at lower redshifts. They also find that sub-mm selected 
galaxies are more dusty per unit stellar mass and more obscured at earlier times, however this second finding is simply based on the trend of averaged
quantities over redshift and thus can be influenced by the IR selection which favours more obscured galaxies as the limiting detectable dust mass increases with redshift.

Before discussing the trends with SFR, mass and redshift, we emphasise that this sample
is not volume or luminosity limited, and there are consequently strong correlations between
these three parameters (with mass and SFR both being higher at higher redshift), simply
as a consequence of the flux-limited sample being explored. While this selection bias
should be borne in mind, none of the following investigation is reliant on using
a volume or luminosity limited sample, and the conclusions are not affected by the flux-limited
sample selection.

The SFR dependence is shown in Figure~\ref{bigfig}(a)-(c). The SFRs used here are
those derived from the H$\alpha$ luminosity, obscuration corrected using the Balmer
decrement and the obscuration curve of \citet{FD:05}, as in \citet{Wij:10}.
It is clear that $L_{IR}$ correlates with the H$\alpha$ derived SFRs, and it is
worth emphasising that systems with the highest SFRs and the highest $L_{IR}/L_{FUV}$ ratios
are pushed to the lowest FUV luminosities. In other words, a UV-selected sample will
always be prone to missing even very high SFR systems with sufficient obscuration.

\begin{figure*}
\centerline{
\includegraphics[width=70mm]{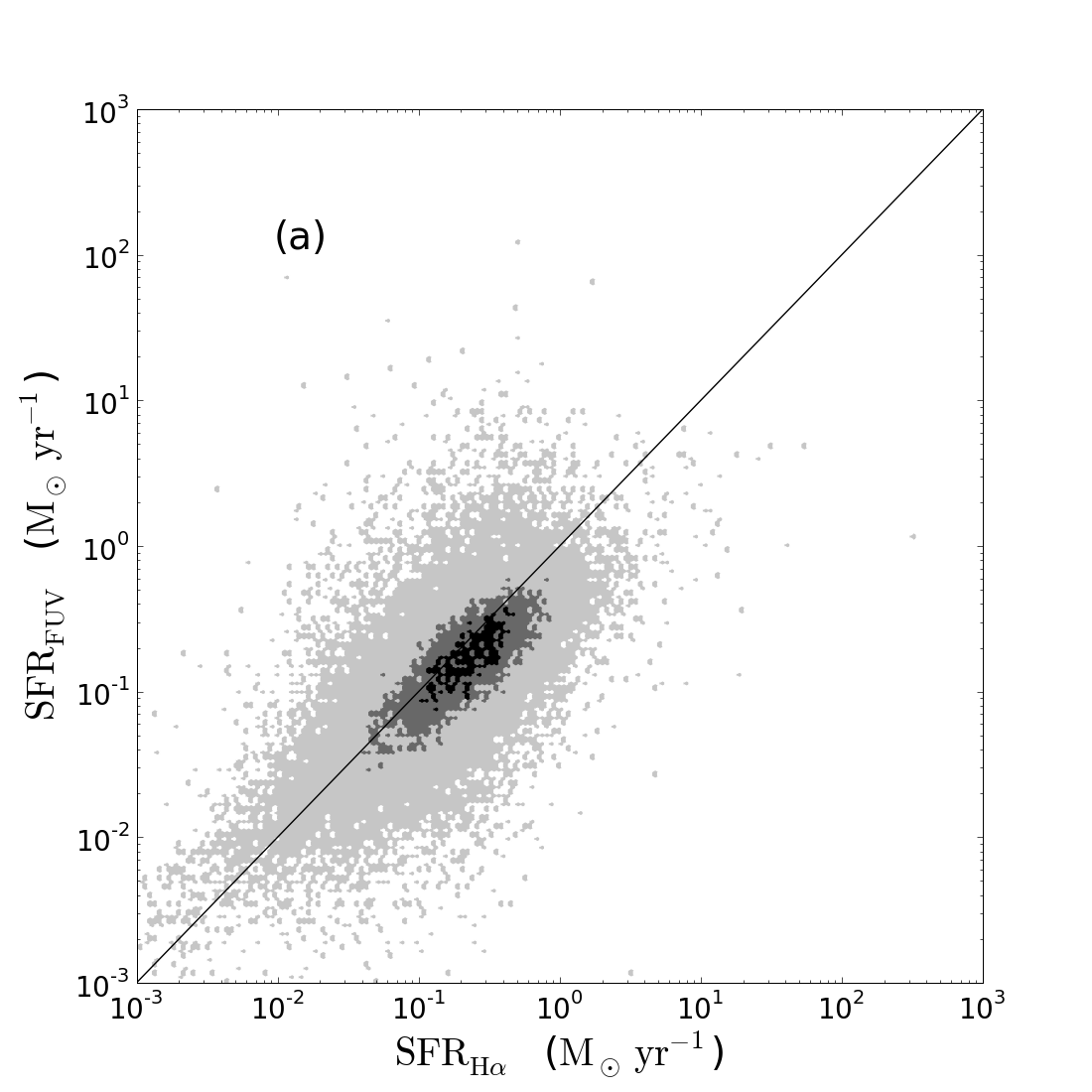}
\includegraphics[width=70mm]{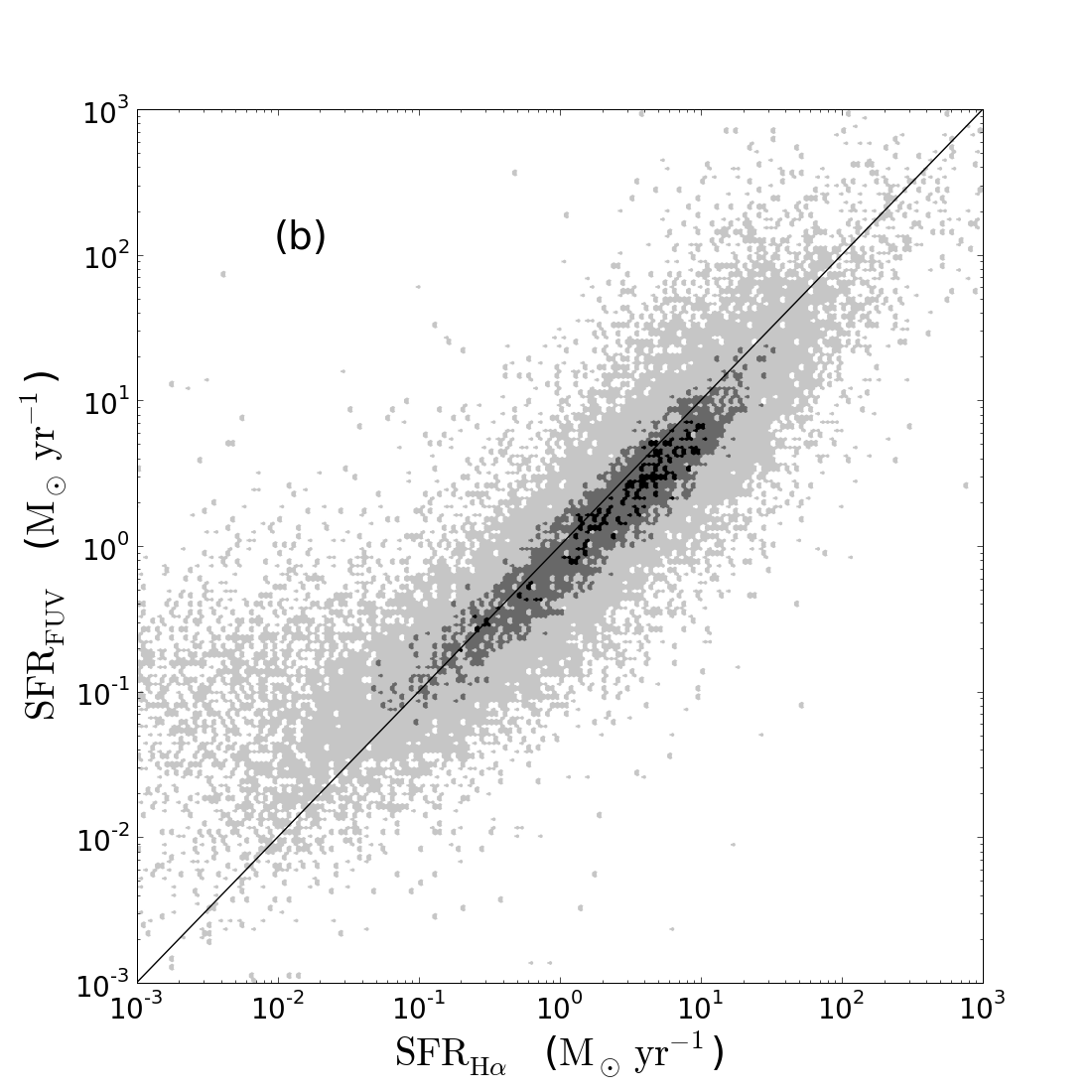}
}
\centerline{\includegraphics[width=70mm]{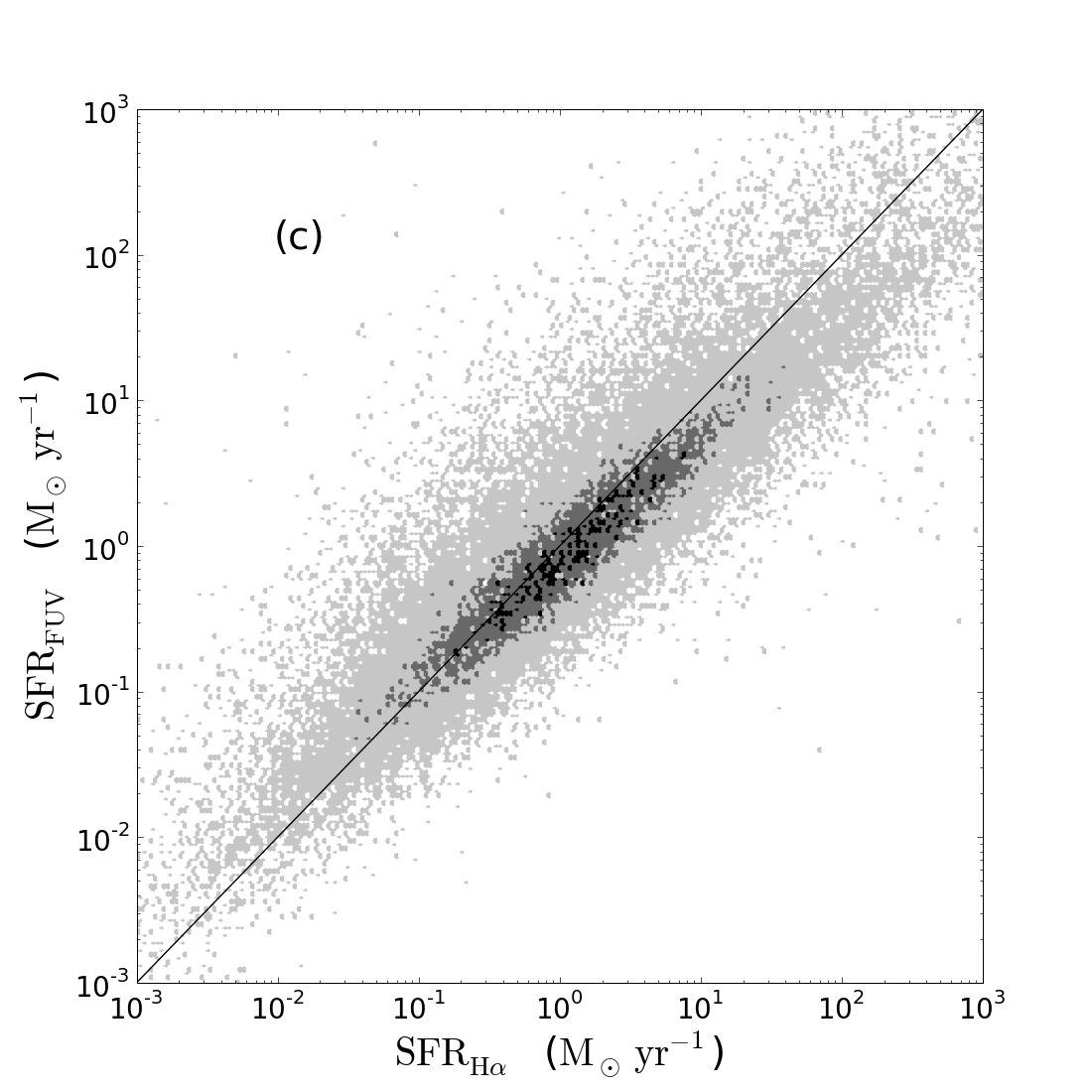}
\includegraphics[width=70mm]{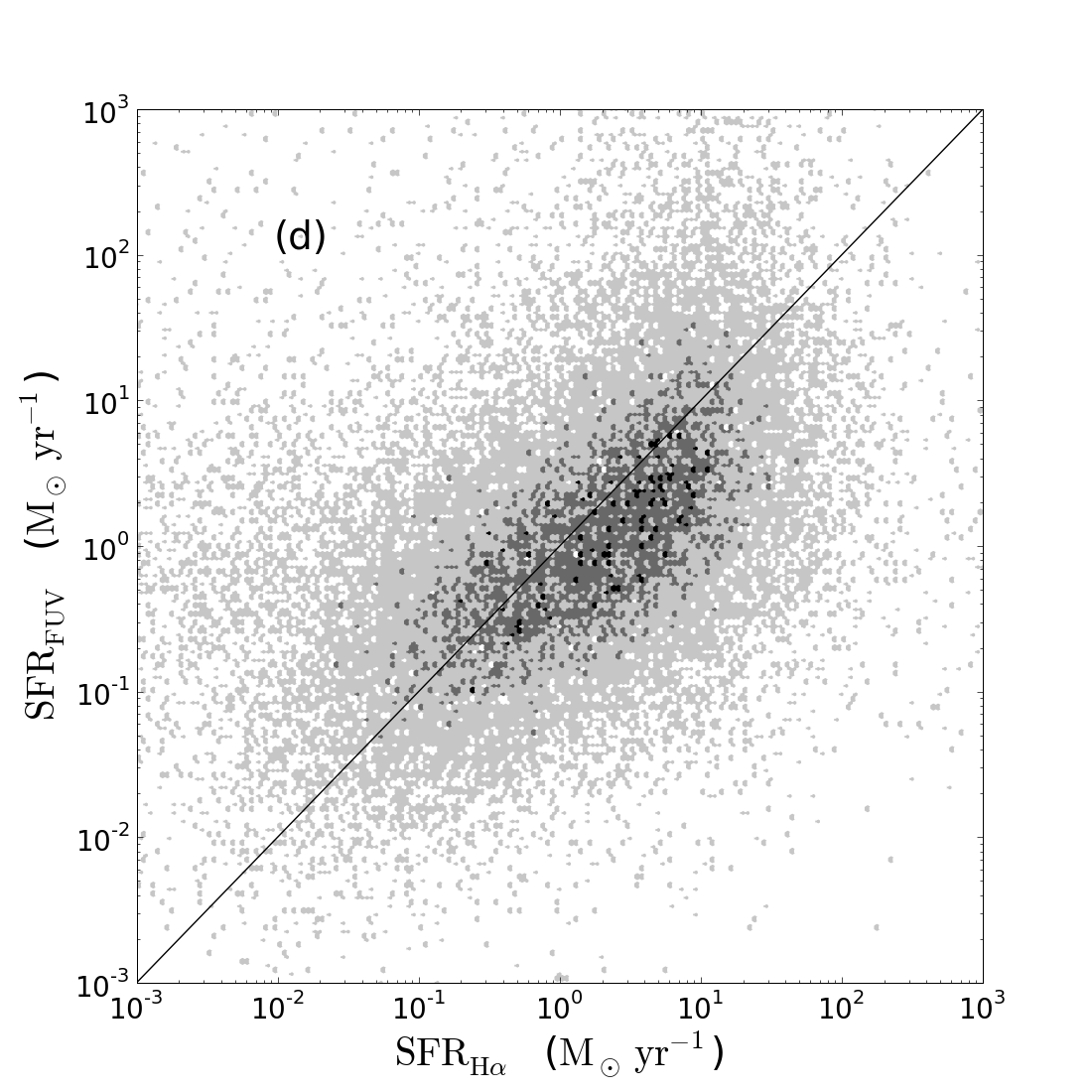}
}
\caption{FUV SFR plotted as a function of H$\alpha$ SFRs using various obscuration corrections.
(a) No obscuration corrections applied. (b) FUV SFR against H$\alpha$ derived SFR,
both obscuration corrected using the Balmer decrement. (c) FUV SFR against H$\alpha$ derived SFR,
both obscuration corrected using $\beta$. (d) FUV SFR corrected for dust using $\beta$
compared against that measured using H$\alpha$, corrected using the Balmer decrement.
The solid line in each panel shows equality between both axes. The figures show the density of points in a given region in the plot where darker the shading the denser the points in that 
region.}
\label{fig:sfrs}
\end{figure*}

The redshift dependence (Figure~\ref{bigfig}(d)-(f)), and the mass dependence
(Figure~\ref{bigfig}(g)-(i)) are both very similar to the SFR dependence. This is
a consequence of both the flux limit (meaning that high-SFR systems tend to be found
at higher redshift) and the relatively tight relationship between stellar mass and SFR.
We can account for this to some degree by looking at the dependence on
specific star formation rates (SSFR), which are sampled more uniformly
with redshift (Figure~\ref{bigfig}(j)-(l)). This shows a very similar trend as for
SFR and mass, with higher luminosity systems having higher SSFRs
in general (a consequence of massive star forming galaxies being able
to support proportionally higher levels of star formation than low mass
systems). There is no strong dependence, though, between $L_{IR}/L_{FUV}$
and SSFR, with a broad range of SSFRs being sampled at any given
value of $L_{IR}/L_{FUV}$.

Figure~\ref{bigfig}(d) echoes Figure~1 of \citet{Buat:10}, but continued to lower
luminosity and redshift, sampling the $0<z<0.35$ range. This demonstrates a clear
trend that continues to lower $L_{IR}/L_{FUV}$ ratios at lower $L_{IR}$ and lower redshift.
The apparent strong correlation between $L_{IR}$ and redshift is a consequence
of the flux-limited sample as well as the strong evolution in the $L_{IR}$ over this redshift range.
\citet{Dye:10} showed that the 250$\mu$m luminosity density increases as $(1 +  z)^{7.1}$ in $0 < z < 0.2$. Similar evolution has also been observed from IRAS and Spitzer/ISO.
\citet{Dun:10} showed that the dust content of galaxies also increases over this redshift range.

It is clear, though, that lower luminosity systems, which in general
show lower $L_{IR}/L_{FUV}$ ratios, tend to have lower SFRs and masses, and are
more easily visible at lower redshift. The tendency for higher luminosity systems (those
with higher SFRs) to have a broader range of obscurations, and on average to be more
heavily obscured, is now well-established \citep[e.g.,][]{Hop:01,Hop:03,Afo:03,Per:03}, and this
tendency is illustrated throughout Figure~\ref{bigfig}.

The distribution of the obscuration measures $\beta$ and Balmer decrement can be
seen in Figure~\ref{bigfig}(m)-(o) and Figure~\ref{bigfig}(p)-(r) respectively. The striking
feature here is the lack of any strong relationship. Reflecting the weak trends seen in
Figure~\ref{bds2}, there is a definite tendency to see systems with
higher values of $\beta$ at higher $L_{IR}/L_{FUV}$ ratios, and these systems are
also those with high $L_{IR}$ and the lowest $L_{FUV}$. The Balmer decrement on
the other hand shows even less of a systematic variation, with only a very weak tendency
toward higher values for high $L_{IR}/L_{FUV}$ and high $L_{IR}$. This again
is likely to be a consequence of the sensitivities of Balmer decrement and $\beta$ to
different optical depths within these galaxies.

The overall conclusions from this exploration are:
\begin{enumerate}
\item We see an obvious relationship between obscuration as measured
by $L_{IR}/L_{FUV}$ and SFR measured using dust corrected H$\alpha$ luminosities. This is consistent with existing
results \citep[e.g.,][]{Hop:03,Afo:03,Per:03}, showing lower obscuration for low SFR
systems, but a large range of obscuration in high SFR systems.
\item We see a similar result with galaxy stellar mass, reflecting the mass-SFR relationship
in galaxies (high SFR systems tend also to be high stellar mass), again with
low mass systems displaying lower obscuration, but high mass galaxies having a broad
range. This is consistent again with the result found by \citet{Buat:09} as a function of $K$-band
luminosity. 
\item The redshift variation is a consequence of the sample selection, with high mass, high SFR
systems only being able to be detected at higher redshift, due to the volume being sampled.
As \citet{Dun:10} shows, there is also likely to be a contribution from evolution in the dust masses in galaxies over this redshift range.
\item The correlation between the $\beta$ parameter and the BD was found to be very weak.
\end{enumerate} 

\section{Star formation rates}
\label{sfrs}
The primary reason for a detailed investigation of the UV spectral slope is in assessing
its utility for making obscuration corrections to UV luminosities, and in many cases to
subsequently calculate SFRs. Here we look at SFRs calculated using FUV and H$\alpha$
luminosities, corrected for obscuration using various combinations of $\beta$ and Balmer
decrement. These SFRs were calculated using the prescription outlined in \citet{Wij:10} 
where the SFR conversion factor was derived using a \citet{BG:03} IMF with a mass range of 
$0.1 M_{\sun}$ to $120 M_{\sun}$ and the population synthesis model by \citet{FR:97}.

\citet{Wij:10} calculated SFRs from FUV luminosities corrected for obscuration using both
Balmer decrement and $\beta$, showing that even when accounting for the offset in
the attenuation from the \citet{Meu:99} prescription, there remains significant scatter in
the relation. This is highlighted again in Figure~\ref{fig:sfrs}. This Figure compares SFRs
derived using dust corrections made using Balmer decrement for both measures,
using $\beta$ for both measures, and using $\beta$ for the FUV and Balmer decrement
for the H$\alpha$ SFRs. The first point to emphasise is that using inconsistent obscuration
measures for different SFR estimates is the primary source of the huge scatter seen.
The scatter in Figure~\ref{fig:sfrs}(d) is a direct reflection
of the scatter between $\beta$ and Balmer decrement (Figure~\ref{bds}).

It is also clear that SFRs calculated using dust obscuration corrections based on the
Balmer decrement give the most self-consistent results (Figure~\ref{bigfig}(b)). When
$\beta$ is used to correct both FUV and H$\alpha$ luminosities a correlation is seen
in panel (c) in Figure~\ref{bigfig}. This is expected as both the FUV and the H$\alpha$ luminosities
share the identical $\beta$ factor leading to the same obscuration factor being applied to both.
Since the uncorrected luminosities show some agreement (Figure~\ref{fig:sfrs}(a)),
multiplying by the same factor merely stretches out the distribution to what is
observed in Figure~\ref{fig:sfrs}(c). The deviation seen at high SFRs, with FUV SFRs tending to be slightly, but
systematically, underestimated compared to H$\alpha$ SFRs, is also visible in the uncorrected luminosities, and reflects
the greater effect of obscuration at UV wavelengths.  This contrasts to the Balmer decrement approach which finds agreement between the two SFR
indicators by using dust corrections appropriate for each wavelength.

Given that $L_{IR}/L_{FUV}$ is (marginally) more tightly correlated with $\beta$ than it is
with Balmer decrement (Figure~\ref{bds2}), it is straightforward to infer that using an
infrared excess to account for obscuration at UV wavelengths will result in similar
effects as seen above with $\beta$. We do not attempt to calculate SFRs directly from $L_{IR}$ in the current investigation, as we are focusing on the 
self-consistency of obscuration corrections between the UV and H$\alpha$ derived SFRs.
A detailed analysis of SFR calibrations from \emph{Herschel} fluxes in multiple passbands is currently underway (da Cunha et al., in prep.). 

Consider the implications of the scatter between $\beta$ and Balmer decrement.
If the origins of the scatter are largely physical, associated with probing different optical depths,
stellar populations and dust geometries, it is perhaps reasonable to argue that $\beta$ is a
more appropriate metric to use in making dust obscuration corrections to UV luminosities,
as the obscuration being probed by $\beta$ is the same as that affecting the luminosity to
be corrected. Similarly, the argument would be made that Balmer decrement would be
the most appropriate obscuration metric to use in correcting H$\alpha$ luminosities. Why,
then does this combination give rise to a poorer comparison between the two estimates
of SFR (Figure~\ref{fig:sfrs}(d)) than using Balmer decrement for both (Figure~\ref{fig:sfrs}(b))?
We explore this further in \S\,\ref{disc} below.

\section{Discussion}
\label{disc}
Given the above limitations, what can we say about the utility of $\beta$ as a proxy for
obscuration in galaxies? First, we emphasise that we are limiting ourselves here
to values of $\beta$ estimated from the two broad UV bands of GALEX. 
More robust constraints on the UV spectral slope \citep{Cal:94} would clearly improve the
situation by removing that contribution to the scatter in the measurement of $\beta$.
\citet{Kong:04} also show that the inclusion of SFR histories also helps to reduce the
scatter when using $\beta$ as an obscuration metric. The star formation rates
of the current sample were investigated in the context of $L_{IR}/L_{FUV}$, and
also against $\beta$, and showed no straightforward correlation. Regression analysis,
using Balmer decrement as the independent variable, and multiple combinations of
$\beta$, SFR, redshift and stellar mass, as the dependent variables, shows little
improvement in the weak correlation already found between Balmer decrement and
$\beta$.

It is also undoubtedly the case that the UV spectral slope is sampling a measure
of obscuration that is physically different from that measured by the Balmer decrement
\citep[e.g.,][]{Cal:97,CF:00}. The use of the Balmer decrement in applying
obscuration corrections to both FUV and H$\alpha$ derived SFRs, however, produces highly
consistent estimates. This suggests that the effect of the potentially different obscurations
on the UV luminosity and the H$\alpha$ luminosity is not a major or systematic effect,
and that the uncertainties associated with the $\beta$ approach are much larger than
these underlying physical differences can explain.

The conclusion we are left with, then, is that, the above limitations associated with
measurement aside, the UV spectral slope is sensitive to many factors, of which obscuration
is only one. These additional factors, including the age of the most recently formed
stellar population, and contributions from older stellar populations, along with
metallicity and IMF slope, are not insignificant, and are challenging to account for in a simple way.
Further investigation of the utility of $\beta$ as an obscuration metric will need to explore
all these effects in a thorough and systematic fashion, ideally with data that samples
the UV spectrum finely.

Finally, we reiterate that the correlation of $L_{IR}/L_{FUV}$ is stronger with the
UV spectral slope than it is with the Balmer decrement, although the correlation coefficients
in both cases are low.

\section{Conclusion}
\label{conc}
We have used a sample of galaxies from the H-ATLAS SDP region, with multiwavelength
photometry and spectroscopy from GAMA to explore the relationships between
the UV spectral slope $\beta$, and the Balmer decrement. We find that there is a
very poor correlation at best between the Balmer decrement and $\beta$, and that
the use of $\beta$ as an obscuration metric suffers from significant limitations.

We see, as found by other authors, a clear but weak dependence between
infrared excess, here estimated using $L_{IR}/L_{FUV}$, and $\beta$. We find at most a
weak trend between $L_{IR}/L_{FUV}$ and Balmer decrement. We also reiterate
the results of \citet{Buat:09} that UV selected samples will be strongly biased against
heavily obscured systems, even of similarly high luminosity to those that enter such samples.
We find consistent results with \citet{Buat:10} regarding the trend of $L_{IR}/L_{FUV}$ with
$L_{IR}$ and redshift, probing in this analysis to lower redshifts and luminosities.
We also see trends with SFR and galaxy mass that reinforce existing correlations and
trends between these properties with luminosity and obscuration.

In summary, we urge caution in the use of $\beta$ as an obscuration metric when it is able to be inferred only from a few broadband photometric measurements, in particular
for systems at high redshift, given the limitations apparent in doing so even for large samples of
well-studied galaxies at low redshift.

\section*{Acknowledgements}
D.B.W.\ acknowledges the support provided by the Denison Scholarship from the School of Physics.
GAMA is a joint European-Australasian project based around a spectroscopic campaign using the
Anglo-Australian Telescope. The GAMA input catalogue is based on data taken from the Sloan Digital
Sky Survey and the UKIRT Infrared Deep Sky Survey. Complementary imaging of the GAMA regions is
being obtained by a number of independent survey programs including GALEX MIS, VST KIDS, VISTA
VIKING, WISE, Herschel-ATLAS, GMRT and ASKAP providing UV to radio coverage. GAMA is funded
by the STFC (UK), the ARC (Australia), the AAO, and the participating institutions. The GAMA website
is http://www.gama-survey.org/ . 
The {\em Herschel}-ATLAS is a project with {\em Herschel}, which is an ESA space observatory
with science
instruments provided by European-led Principal Investigator consortia and with important
participation from NASA. The H-ATLAS website is http://www.h-atlas.org/ .

\label{lastpage}

\end{document}